\documentclass[twocolumn,preprintnumbers,amsmath,amssymb]{revtex4-1}
\usepackage{graphicx}
\usepackage{dcolumn}
\usepackage{bm}
\begin{document}

\title{Frequency analysis and dynamic aperture studies in ELENA with realistic 3D magnetic fields  }
\author{Lajos Bojt\'ar}
\email{ Lajos.Bojtar@cern.ch }
\affiliation{ CERN, CH-1211 Gen\`eve 23\\}
\date{\today}

\begin{abstract}
We briefly present recent progress with our algorithm and its implementation called SIMPA described in a previous paper~\cite{BOJTAR2019162841}. The algorithm has a new and unique approach to long-term 4D tracking of charged particles in arbitrary static electromagnetic fields. Using the improvements described in this paper, we made frequency analysis and dynamic aperture studies in ELENA. The effect of the end fields and the perturbation introduced by the magnetic system of the electron cooler on dynamic aperture is shown. A special feature of this study is that we have not introduced any multipole errors into the model. The dynamic aperture calculated in this paper is the direct consequence of the geometry of the magnetic elements. Based on the results, we make a few suggestions to reduce the losses during the deceleration of the beam.
\end{abstract}

\maketitle  
 
\section{ Introduction }

The importance of fringe fields in small rings is well known and it has been taken into account for multipole magnets at various degrees for decades   \cite{LEEWHITING1969305,PhysRevSTAB.3.124001,FOREST1988474,PhysRevSTAB.18.064001}. The approach to particle tracking we described in~\cite{BOJTAR2019162841} naturally includes the end fields altogether and for all kinds of elements with the same treatment.  The aim of this paper is two-fold. One is to apply our algorithm described in~\cite{BOJTAR2019162841} on the Extra Low ENergy Antiproton (ELENA) ring~\cite{ElenaDesignRep}, the 30.4 meters circumference machine built at CERN to decelerate antiprotons. Another aim is to gain insight into the beam dynamics of ELENA by frequency analysis and dynamics aperture studies.  

Long-term tracking of charged particles is a fundamental problem of accelerator physics, plasma physics and it is important in astrophysics. In a recent paper~\cite{BOJTAR2019162841} we described a new algorithm allowing long-term symplectic integration of charged particle trajectories in arbitrary static magnetic and electric fields. We introduced several new ideas and pointed out the importance of two requirements for a physically valid description of the electromagnetic fields. 

One important point is that the potentials describing static magnetic and electric fields must obey the Laplace equation everywhere in the domain of interest. This is a necessary condition to get physically valid fields from the potentials. Failing to satisfy this condition leads to spurious sources~\cite{BRACKBILL1980426} which in turn leads to an energy drift during the tracking. All 3D grid methods suffer from this problem because there is no known interpolation satisfying the Laplace equation while being continuous between the 3D grid cells.

In  accelerator physics, transfer map methods~\cite{PhysRevSTAB.13.064001,PhysRevSTAB.9.052001,VENTURINI1999387} are also applicable to handle realistic 3D electromagnetic fields. These are conceptually rather different and they assume the existence of a closed orbit, which puts restrictions to the range of problems it can handle. FFAs, stellators, tokamaks for example are difficult to handle with transfer maps, but our algorithm can treat those cases without complications. 

Another approach to deal with the problem of the representation of 3D fields is to take the Fourier transform of the field in three variables~\cite{SLAC_Chang, epac2006_Bahrdt, Titze2016, doi:10.1029/2005JA011382}. 
One disadvantage of this method is the high number of harmonics needed when the field has detailed local features.

The second requirement for a physically valid description of the field is the continuity of the potentials. Cutting the potentials at some distance, even far from a magnetic element, leads to a systematic error which accumulates during the tracking leading to an energy drift and non-preservation of the phase space volume. This has been recognized also by others~\cite{DragtBook2018,PhysRevSTAB.3.124001,PhysRevSTAB.18.064001} and mitigations  has been used to handle this effect. Our method avoids cutting the potentials by treating the entire volume of interest as a whole.
 
Before this study, little was known about the effect of the magnetic system of the electron cooler on the ELENA beam dynamics. More specifically, the effect of the toroid coils in the cooler was mostly unknown, not due to lack of effort, but due to the lack of appropriate tracking tools. A considerable amount of work went into the development of a tracking tool by colleges at CERN to address this issue \cite{Priv_Comm_Beloshit}. The tool was based on a field map given on a 3D grid. It produced manifestly non-symplectic behaviour already in a few hundred turns. 

This triggered the author to devise the algorithm ~\cite{BOJTAR2019162841} and implement it. The name of the software is SIMPA, an abreviation of \underline{s}ymplectic \underline{i}ntegration through \underline{m}ono\underline{p}ole \underline{a}rrangements. At present it is a Java library. A lot of work remains to be done before it can be released for general use. It is intended to become a  tool to produce field maps suitable for long-term, charged particle tracking in general, not only in the domain of accelerator physics. 
It  contains  code to use the generated field maps in actual trackings. Recent progress \cite{PhysRevE.94.043303} made possible to implement an explicit second order symplectic integrator for non-separable Hamiltonians, which arise when magnetic fields are involved.

The simulation tools needed for frequency analysis and  dynamics aperture studies described in this paper were written  in Java using the SIMPA library. Both the simulation source code and the library source code is available from the author upon request. The SIMPA source code will be released under an open source license in the future. People interested in the code are encouraged to contact the author. 

The algorithm described in~\cite{BOJTAR2019162841} was applied to a simple test machine with a beam region having a constant circular profile. This was covered with a single line of overlapping spheres. See Fig. 4. in~\cite{BOJTAR2019162841} for an illustration. Actual machines like ELENA have beam regions with more complex geometries. Much progress has been made to extend the algorithm to general volumes. Full description of the enhanced algorithm is outside the scope of this work and will be described separately, only a summary follows.

The paper is organized in the following way. We present the progress made to cover general volumes and explain also how  this leads to gain in speed and accuracy. Then a number of basic checks are presented, comparing the model of ELENA built with SIMPA to the MAD-X ~\cite{madx} model. In the frequency analysis section the effect of fringe fields and the magnetic system of the electron cooler is investigated. The dynamic aperture section quantifies the effect of resonance lines identified as relevant by the frequency analysis. In the conclusions we suggest a few possible improvements for the performance of ELENA.

\section{ The tracking algorithm }
\subsection{ A short summary  }

We recommend reading the previous paper~\cite{BOJTAR2019162841}  to understand the algorithm in detail, as only a summary is provided here.

Symplectic integrators keep the error resulting from the integration itself bounded, but can not cure the errors coming from the representation of the fields. These are two separate sources of errors. It is crucial to have a physically valid representation of the fields. When an electromagnetic field not satisfying the Maxwell equations is given to a symplectic integrator the integration will be still symplectic, but the result will describe something other than the true physical behavior of the system. A common example is the energy drift due to spurious sources appearing in 3D grid methods ~\cite{BRACKBILL1980426}.

The potentials describing the fields must satisfy the Laplace equation and should be continuous everywhere in the beam region to obtain accurate results in long-term trackings. To comply with these conditions, the potentials are expressed analytically in terms of their sources. We use hypothetical magnetic and electric monopoles and current loops as sources. These sources are placed outside of the volume of interest, at some distance from the boundary, and their strength is set such that they reproduce the normal component of the magnetic or electric field at the boundary by solving a system of linear equations. 

The normal component of a magnetic or electric field determines completely the field inside a closed simply-connected volume. For a multiply-connected volume, this is not always true. If the integral of the magnetic field along a closed curve is not zero, then magnetic monopoles alone can not reproduce the magnetic field. One example of this situation is the solenoid magnet. In these cases, the non-conservative part of the vector potential must be provided by one or more current loops. The word non-conservative should be understood here in a mathematical sense, the vector field having a  non-zero integral along a closed loop , that is $\oint \mathbf{v} \cdot d{\mathbf{r}} \neq 0$.

After the potentials are reproduced at the boundary by the sources, they can be evaluated anywhere inside the volume analytically. The potentials satisfy the Laplace equation and they are continuous. However, this method is too slow to be practical. We will refer to this part of the algorithm as slow evaluation. We need a much faster calculation of the potentials to be of practical use.

Several orders of magnitude improvement can be achieved by using a local description of the potentials.  Spherical harmonics scaled appropriately are called solid harmonics. Regular solid harmonics are the canonical representation for harmonic functions inside a sphere ~\cite{SHUnitSphere}. A key characteristic of the algorithm is the description of vector and scalar potentials by solid harmonics inside a set of overlapping spheres covering the volume of interest.

The potentials satisfy exactly the Laplace equation inside the spheres. The discontinuity between the spheres decrease exponentially with the degree of solid harmonics expansion and can be easily kept close to machine precision. The representation of the potentials in terms of solid harmonics is optimal in terms of memory and allows fast evaluation. We will refer to this part of the algorithm as fast evaluation. The speed of the evaluation is strongly dependent on the degree of the solid harmonics expansion. This property is the base for the speed improvements we will describe later.

One might ask what is the point of having a slow evaluation when we have a fast one. To calculate the solid harmonics coefficients for each ball in the fast evaluation we need the values of the scalar and vector potentials. These potentials must be continuous and must satisfy the Laplace equation. This is the role of the slow evaluation. CAD software or measurements can not provide the potentials satisfying these conditions.   

\subsection{ Extension to general domains }

In ~\cite{BOJTAR2019162841} we show cased the algorithm on a simple accelerator. The beam region had a circular shape everywhere with a constant 3 cm radius all along the reference orbit. This was covered by a single row of overlapping spheres with a 4 cm radius. This solution to cover the beam region is sufficient to demonstrate the method, but real accelerators have a more complex aperture in general. Apart from that, our algorithm aims to apply to a wide range of problems, not only in the domain of particle accelerators. 

The usual way to describe a general volume is to give its boundary surface by triangulation. A common format to describe a triangulated surface is the Standard Tessellation Language abbreviated as STL. The boundary of the volume of interest is described by an STL file in SIMPA. There are many CAD software able to produce STL files, however, to construct the aperture for an accelerator using CAD software is not a trivial task. There is a utility implemented in SIMPA facilitating the construction of the STL file. It takes the design orbit and a list of transfer profiles as the input and extrudes those profiles along the design orbit. For each transfer profile, the user must specify the longitudinal position where the extrusion starts and ends along the orbit. The transverse profiles are described as polygons. The number of segments for each polygon is restricted to the same value, making the implementation of the triangulation straightforward.

Once the boundary of the beam region is specified by an STL file the next step is to fill it with overlapping spheres, such that the entire beam region is covered without gaps. There are constraints concerning the size of the spheres. The slow evaluation calculates the potentials from sources, which are placed at some distance from the boundary. The spheres should be small enough to not overlap with the sources. The solid harmonics expansion works only if there are no sources inside the spheres.

It is practical to place the center of the spheres onto a regular lattice. This simplifies the construction of the cover as well as finding  the spheres when the potentials are evaluated. There are many possible lattices, but only the hexagonal close-packed ( HCP ) and the face-centered cubic (FCC) lattices have the optimal packing density, about 0.74. We chose the HCP lattice. The coordinates of the sphere centers on an infinite HCP lattice can be obtained by three simple expressions. We keep only those spheres from the infinite lattice which are necessary to cover the beam region.

The first step is to find the indices of one initial sphere on the lattice, which has its center inside the beam region. Then a 3D version of an algorithm from computer graphics, called flood-fill, is applied to collect all sphere centers inside the boundary. In essence, the flood-fill algorithm checks neighbouring lattice locations recursively, if they are inside or outside the boundary. After this step, we have a list of spheres which covers most of the beam region, but not all. Near the boundary, there can be gaps.

\begin{figure}
\includegraphics [scale=0.26]{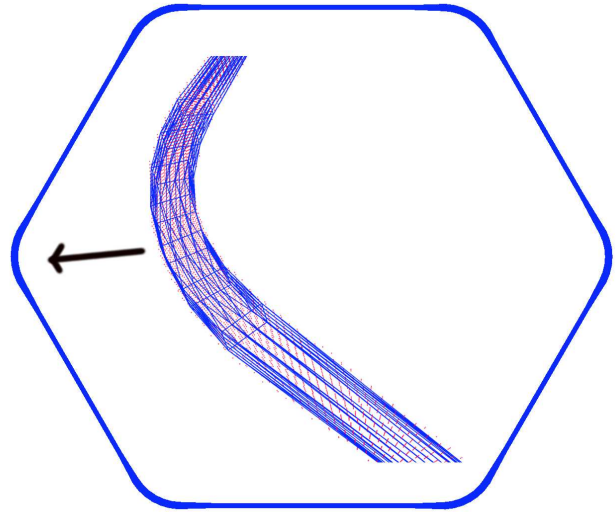} 
\caption{ \label{fig:elenaring} Triangulated boundary of the ELENA beam region. The interior plot is an expanded 3D view of an arc. The red dots are the centers of the covering spheres located on an HCP lattice. }
\end{figure}

A second step is necessary to cover the beam region fully. In the second phase, all spheres kept in the first step are checked in the following way. If the sphere in question has missing neighbours in the lattice, then we take the half-line from the sphere center to the center of the missing neighbour and calculate its intersection point with the sphere in question. If this point is inside the boundary of the beam region, then we add the missing sphere to the cover. We do this for all missing neighbours, hence eliminating gaps. We should note this second step can be applied only if the distance of sources providing the potentials are further from the boundary than the diameter of the covering spheres. In the present study that is the case. Otherwise, some spheres would have sources inside, which is not allowed. A more sophisticated second step of the covering process could be applied by removing the restriction to put sphere centers only to the lattice points and allowing variable sphere sizes in the second step. This is a non-trivial problem in computational geometry, but certainly has a satisfactory solution. FIG. \ref{fig:elenaring} shows the ELENA beam region and the centers of the covering spheres. The radius of the spheres was 1 cm in this study and 75951 of them were needed to cover the aperture.

\subsection{Improvements in speed and accuracy}
Covering the volume of interest as described above makes the method applicable to a much wider range of problems, but this is not the only advantage. In general, the error of the spherical harmonic approximation on the surface of a sphere is less than the first term omitted from the infinite sum of spherical harmonics times a constant \cite{SHUnitSphere}. Let us inspect Eq. (9) in~\cite{BOJTAR2019162841}, which is restated here for convenience.
\begin{equation}
f(r,\theta,\phi) = \frac{1}{R} \, \sum_{\ell=0}^{\ell _{\textrm{max}}} \, \sum_{m=-\ell}^{\ell}\, r^\ell  \, c_{\ell m} \, \tilde{P}_\ell ^m (\cos\theta) \, \Phi(\phi;m).  \label{eq_pot_value}
\end{equation} 
In this equation the value of $r$ is between zero and $R$, the radius of the sphere. We can set $r=R$ and calculate the maximum error of the solid harmonics approximation as $ \epsilon R^{\ell _{\textrm{max}}} $, where $\epsilon$ is the error due to the first omitted term in the sum with degree $ \ell _{\textrm{max}}+1$. Because of the presence of the term $r^{\ell}$, the biggest possible error in the approximation strongly depends both on $R$ and $\ell _{\textrm{max}}$.
 
For instance, in~\cite{BOJTAR2019162841}, we used a single row cover with $R=4$ cm and $\ell _{\textrm{max}}=38$. In this study $R=1$ cm. If we had used the same $\ell _{\textrm{max}}=38$ with $R=1$ cm spheres, the maximum error of the approximation   $ \epsilon R^{\ell _{\textrm{max}}}$ would have been a factor $4^{38}$ smaller, assuming infinite precision arithmetic and the two spheres having the same center. This is a big number, so we can reduce $\ell _{\textrm{max}}$ and the error of the solid harmonics approximation still stays close to machine precision.
 
 In this study we used $\ell _{\textrm{max}}=12$ uniformly for all spheres. 
 This speeds up the evaluation of the potentials and their derivatives significantly, because the number of operations to calculate the sum in  Eq. (\ref{eq_pot_value})  is proportional to  ${\ell^2_{\textrm{max}}}$.
 On Intel i7-6700 CPU at 3.40 GHz, using a single core with hyper-threading enabled, we got $2.7 \times 10^5$ evaluations per second of the vector potential and its derivatives with $\ell _{\textrm{max}}=12$. This should be compared against $5.8 \times 10^4$  evaluations with the single row cover and $\ell _{\textrm{max}}=38$,  a factor 4.66 improvement. 
 
 Further speed improvement can be achieved by changing  $\ell _{\textrm{max}}$ between the spheres to keep the error of the approximation below a prescribed value. There is also ample space for optimization by improving memory locality, better use of vector instructions of the CPU, and parallelization. This will be implemented in the future. 

There is one downside of the reduced sphere diameter, that is the increased memory use. The number of spheres needed to cover a given volume is proportional to 1/${R^3}$. The number of coefficients in a solid harmonics approximation is proportional to  ${\ell^2_{\textrm{max}}}$, which in turn depends on $R$. Together, the number of coefficients needed to describe a potential in a given volume with some  prescribed accuracy is $n= c \  {\ell^2_{\textrm{max}}} /R^3 $, where $c$ is a constant. The constant $c$ depends on the volume to be covered. In the present study the size of the field map is 448 MB. Using a single row cover with  $\ell _{\textrm{max}}=38$ it is only 29 MB. Memory is traded for speed.
\section{ Preparations }
\subsection{Preparing the field maps}
As a first step, the field of each type of magnet in ELENA has to be expressed as a collection of magnetic monopoles or a combination of magnetic monopoles and current loops. This provides continuous and analytic potential everywhere in the beam region. To do so, the magnetic field values have been obtained from the CAD software OPERA for each magnet at specific points on a surface surrounding the beam region. This surface is close to the poles of the magnet. The magnetic monopoles are placed outside of the surface at 2.5 cm distance. Then a system of linear equation was solved to find the strengths of the monopoles for each magnet type. The relative precision of the reproduction of the magnetic field from the vector potential is typically between $10^{-3}$ and $10^{-4}$.

Once each type of magnets in the machine has been expressed as a collection of monopoles and current loops, the next step is to assemble the ELENA ring from these collections. This has been done by rotating and translating the collections of sources to the correct place. The entry and exit coordinates of the magnets have been obtained from the MAD-X model with the survey command.

After the machine is assembled, a particle can be tracked to check if the procedure used so far has been correct. This tracking is slow because the vector potential of all sources has to be evaluated at each time step. Nevertheless, it is fast enough to track a few 10's of turns to calculate the closed orbit and set the values of the orbit correctors, which is necessary when the electron cooler is included. 

The magnets of the ELENA ring have been organized into magnet groups. Usually, a magnet group consists of magnets connected to the same power supply. For each magnet group, a field map was produced. These field maps consist of a large number of overlapping spheres, each with some number of coefficients determined by the maximum degree of solid harmonic expansion as described earlier. The location and size of the spheres are calculated once and it is common for all field maps. This makes it easy to scale field maps individually and sum them into a single map before tracking.  
\subsection{Basic checks}
It is reassuring to implement some checks and compare the results to an established tool when this is possible. Since there is already a MADX-PTC model available for ELENA, we use this model as a reference for comparison. There are several global parameters of the machine which can serve as cross-checks, such as the tunes and the optical functions. We compared the optical $\beta$ and dispersion functions to those obtained with MAD-X at the same currents for the quadrupoles. In this comparison, the electron cooler magnets and the solenoid compensators were turned off in both models as well as all orbit corrector magnets.

\begin{figure}
\includegraphics [scale=0.72]{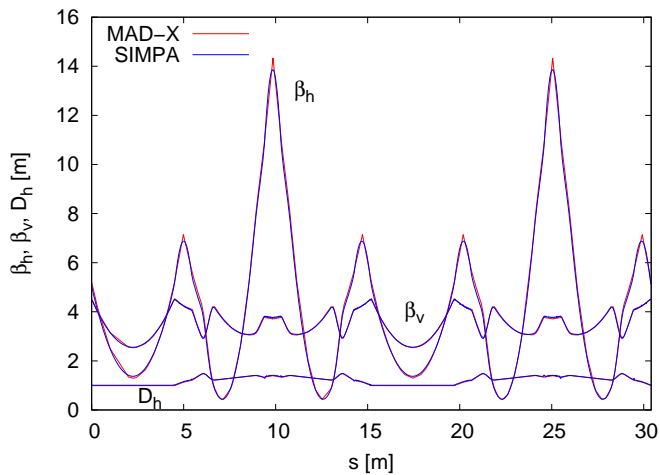}
\caption{\label{fig:twiss} Comparison of the Twiss functions $\beta_h, \beta_v$ and $D_h$  calculated  by SIMPA and MAD-X at the working point $Q_h=2.454$ and $Q_v=1.416$. The functions are hardly distinguishable on the plot.}
\end{figure}

In our model, the optical functions are obtained from tracking. This might look more complicated than the usual linear optics model, but it is necessary because we deal with arbitrary 3D fields. To obtain the Twiss functions, we tracked a $\overline{p}$ with nominal energy while saving the phase space variables at each time step.  The symplectic integration method we implemented according to \cite{PhysRevE.94.043303} uses canonical variables, but these were converted to positions and angles. The procedure results in many Poincar\'{e} maps along the orbit. We then fitted an ellipse to the phase space variables at each longitudinal position. From the parameters of the ellipse, the Twiss functions were calculated by well-known formulas. To get the dispersion, another particle with a different $dp/p$ was tracked and the center of the fitted ellipses was compared to the $dp/p=0$ case. The $\beta$ and $D$  functions are plotted in FIG. \ref{fig:twiss}. The agreement with MAD-X is remarkable if one considers the completely different calculation of the Twiss functions.

The tunes were also compared. We chose the working point $Q_h=2.454$ and 
$Q_v=1.416$. The reason for this choice will be clear later in light of the frequency analysis results. The tunes calculated by MAD-X with the  PTC ~\cite{PTC2002} Twiss command are reasonably close, $Q_h=2.479$, and  $Q_v=1.42$.  

A $\overline{p}$ with $dp/p =0.001$ was tracked and compared to the $\overline{p}$ with nominal energy to calculate the linear chromaticity. The tunes for both $\overline{p}$ were calculated from the tracking results with a very precise frequency analysis tool called PyNAFF \citep{pynaff}. The chromaticities we got are $\xi_h=-4.72$ and $\xi_v=-0.15$, which is significantly different from the results obtained with MADX-PTC, $\xi_h=-3.3$ and $\xi_v=-1.81$. One possible explanation for this discrepancy is that the end fields in MADX-PTC are taken into account by the FINT and HGAP parameters~\cite{madx,Brown:283218} for the bending magnets, while our algorithm handles the end fields nearly exactly for any element. We have not made measurements during commissioning with similar conditions in ELENA.
\section{Frequency Analysis}
\subsection{Motivation}
Small machines have  special challenges, which can be neglected in large rings. One of these is the increased influence of fringe fields on the beam dynamics. ELENA  is a ring with 30.4 meters circumference. It is mandatory to take into account the effect of fringe fields already during the design of machines of this size. The ELENA MADX-PTC model includes the fringe fields  for the main bending magnets, which is one of the main sources of nonlinearities. Another important source of nonlinearities is the magnetic system of the electron cooler, but it is not included in the  MADX-PTC model. Our model includes all fringe fields and also the magnetic field of the electron cooler  with high fidelity.

\subsection{The method}
According to the KAM theory, invariant tori of a conservative dynamical system are stable under small perturbations. These tori describe quasi-periodic orbits in phase space. The motion of the system can be described by action-angle variables. On invariant tori the action variables are constant and the angle variables have fixed irrational and Diophantine frequencies.  When the frequencies are rational numbers, they are related to each other by integers and resonance occurs. The condition for resonance in a particle accelerator in 4D phase space is given as 
\begin{equation}
 n \nu_x + m \nu_y   = p  \label{resonanceCond}
\end{equation} 
,where $\nu_x, \nu_y$ are the horizontal and vertical tunes and $n,m,p$ are integers, not all zero. A more detailed accessible explanation can be found in \cite{Scandale:1994ci} for instance. Although the frequencies where a resonance occurs are known a priory, their strength or significance is not known. One can track particles to investigate the strength of the resonances. Doing so for a large number of turns is time-consuming. Fortunately, there are some early indicators to detect chaotic behavior. One early indicator is the frequency shift.

An early use of frequency analysis for conservative Hamiltonian systems by Laskar \cite{Laskar88} was to investigate the stability of the solar system. Later it was applied to other domains among them accelerator physics~\cite{Bartolini1996_TuneEval,Bartolini1996_PreciseTune}. For a comprehensive description of the method see \cite{Laskar93}. The precision of Laskar's NAFF~\cite{LASKAR1990266} algorithm is the feature that makes it interesting. While the frequency error of the standard FFT is proportional to $1/N$, the error of NAFF is proportional to  $1/N^3$, where $N$ is the number of samples. This allows us to determine the characteristic frequencies of the motion with a  small number of turns.

The first characteristic frequencies for each plane of the motion are the tunes. The drift in the tunes can serve as an early indicator of the long-term stability of the motion \cite{Todesco96_FastIndicators}. For initial conditions corresponding to chaotic trajectories, the frequency can only be defined for a given time interval. Calculating this local frequency for two consecutive time intervals and taking their difference we can calculate tune shifts $\Delta \nu_{x,y}$.  For a coasting beam, the figure of merit can be defined as:
\begin{equation}
  D= log_{10}(\sqrt{\Delta \nu_x ^2 + \Delta \nu_y ^2}) \label{diffCoeff}
\end{equation} 
$D$ calculated with this definition can be used to identify stable areas in the tune diagram.

To identify promising working points, we scanned a large range of tunes in the horizontal and vertical planes and for each initial condition the figure of merit $D$ was calculated. The ELENA machine has three quadrupole families, which give three parameters for the tune scan. There is also a constraint for the dispersion in the electron cooler region, which is fixed to 1 meter. This constraint selects a 2D plane in the 3D space of possible quadrupole currents. The two-dimensional tune space was scanned between $2.2 - 2.52$  and $1.2 - 1.5$ horizontally and vertically by changing the currents in the three quadrupole families. The quadrupole currents from the estimated tunes were calculated by a second-order polynomial. This polynomial is identical to the one used in the ELENA control system. The estimated tunes are a bit different from the tunes obtained from the tracking results by PyNAFF, which we call 'numerically measured tune'. This is because the second order polynomial mapping estimated tunes to quadrupole currents is only an approximation based on linear optics. 

The ranges of tunes in both directions are large, and the optics can change significantly during the scan. The initial conditions have to be varied to compensate for the optics change and keep the emittances approximately constant. To achieve this, we first calculated the optics at several points free of low order resonance lines in the 2D tune space. The optical functions were interpolated from those points, then the initial conditions of the particle were calculated such that the single-particle emittance stay approximately 30 $[\pi \  mm \ mrad]$ in both planes. This particular value for the emittance was chosen to be large enough to see clearly the non-linear effects due to the fringe fields and the electron cooler, but without having too much losses due to the physical aperture of the machine. In this paper emittances and acceptances are defined as the area of the ellipse corresponding to the single particle Courant-Snyder invariant, unless stated otherwise.

The tunes were scanned in 160 steps in both directions giving  25600 initial conditions. Each particle was tracked for 300 turns and the phase space variables were saved at a single longitudinal position for each turn. The tracking result was post-processed with PyNAFF, a Python implementation of the NAFF algorithm~\cite{LASKAR1990266}. To calculate $\Delta \nu_{x,y}$, we split the 300 turns into two sets. Then the figure of merit $D$ in Eq. (\ref{diffCoeff}) was calculated for each point in the diagram and plotted as colors. 
\subsection{Results}
\begin{figure}  
\includegraphics[scale=0.49]{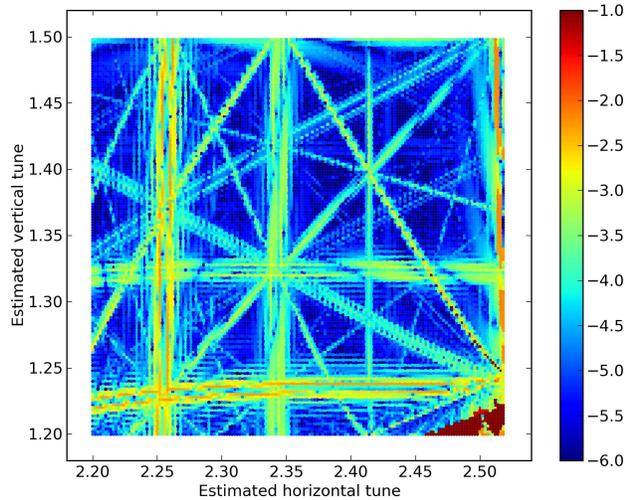}
\caption{\label{fig:scanEst} $D$ plotted as colors against the estimated tunes for the bare ELENA machine, consisting of only the main bendings and the three quadrupole families. The red dots correspond to lost particles.}   
\end{figure}

FIG. \ref{fig:scanEst} shows the figure of merit $D$  for the bare ELENA machine with bendings and quadrupoles only, plotted against the estimated tunes as colors. The term `estimated tune' means the input for the polynomial, mapping the desired tunes to quadrupole currents, which is slightly different from the actual tunes. The red dots indicate the initial conditions which resulted in particle loss in less than 300 turns.

\begin{figure} 
\includegraphics[scale=0.49]{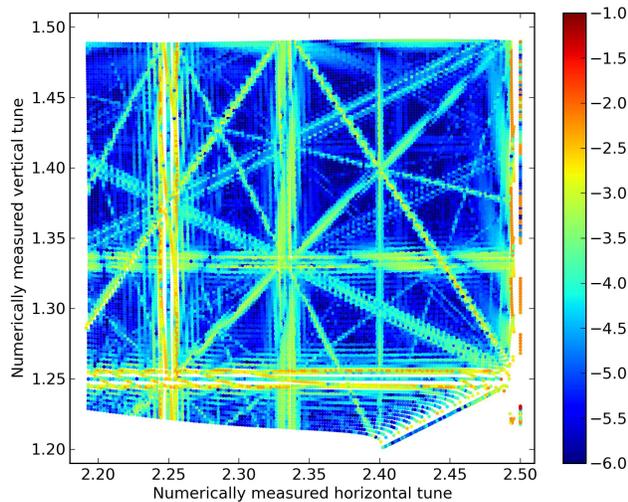}
\caption{\label{fig:scan} $D$ plotted as colors against the numerically measured tunes for the bare ELENA machine, consisting of only the main bendings and the three quadrupole families.} 
\end{figure}

Instead of the estimated tunes, we can plot $D$ against the tunes calculated by pyNAFF as shown in FIG. \ref{fig:scan}. This gives a precise value for the tunes, but this tune can not be calculated for the lost particles, that is why FIG. \ref{fig:scan} has no red dots inside. FIG. \ref{fig:scanEst} and  FIG. \ref{fig:scan} are somewhat complementary.

On both plots, many higher-order resonance lines can be seen. It should be emphasized, these resonance lines are present also in a perfectly manufactured, built, and aligned machine. We have not put any imperfections into the model. All the resonance lines are the results of the fringe fields of the bending magnets and the quadrupoles. They are the direct consequence of the geometry of the magnets.

The next step was to repeat the exercise with the magnetic elements of the electron cooler included. Those are the main solenoid coil, two toroid coils at the ends, two solenoid compensator magnets, and several orbit correctors including the two `Kyoto type'. We did all trackings at the lowest energy of 100 KeV, because the effect of the electron cooler is the biggest there. The cooler has fixed currents for all coils during the deceleration cycle.

\begin{figure}
\includegraphics[scale=0.49]{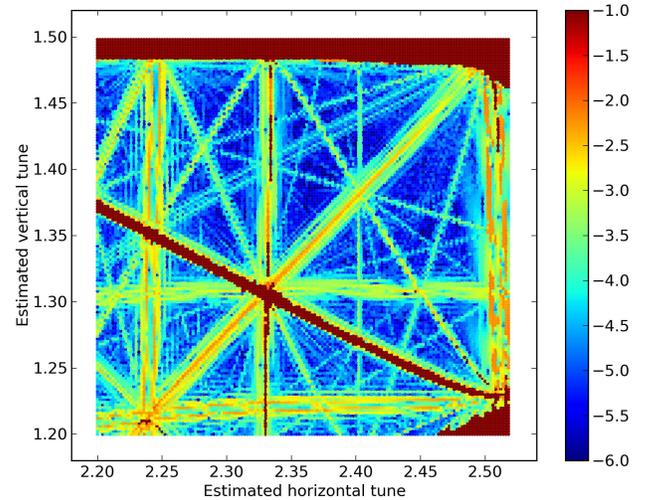}
\caption{\label{fig:scanECEst} $D$ plotted as colors against the estimated tunes for the ELENA machine with electron cooler. The red dots correspond to lost particles.} 
\end{figure}

\begin{figure}
\includegraphics[scale=0.49]{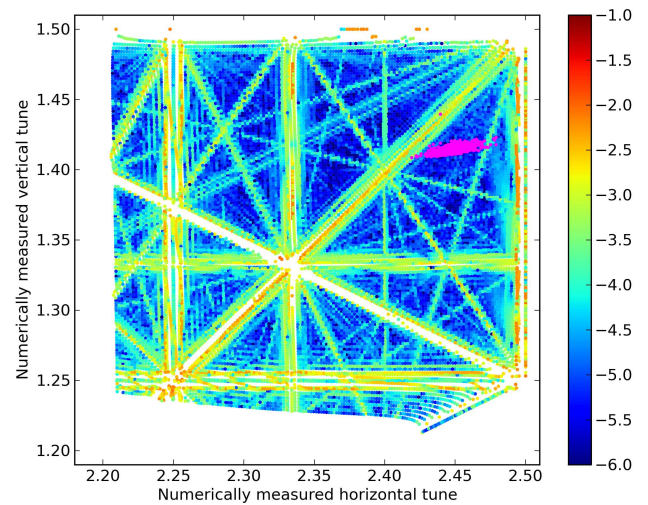}
\caption{\label{fig:scanEC} $D$ plotted as colors against the numerically measured tunes for the ELENA machine with electron cooler. The magenta dots in this plot correspond to the tunes of particles which survived $10^4$ turns in a bunch filling the acceptance, without space charge included in the model.} 
\end{figure}

The step size was set to 1 cm while tracking the bare machine. With the electron cooler included, it was reduced to 0.5 cm. This was needed because the cooler introduced some spatially fast-changing magnetic fields.
FIG. \ref{fig:scanECEst} shows $D$ against the estimated tunes and FIG. \ref{fig:scanEC} against the numerically measured tunes. It is immediately apparent that the magnets of the electron cooler have a significant effect on the beam dynamics. Many resonance lines became stronger and wider and particle losses are more frequent at the strongest resonance lines.

It is natural to ask, how much space is available for the beam in these diagrams. To answer this question we sampled 1000 particles from a Gaussian distribution with $2 \ \sigma$ RMS emittances  $75 \  [\pi \   mm \ mrad]$ in both planes, which is the  acceptance of the machine and $dp/p = \pm 9 \ 10^{-3} $ which is about the momentum acceptance given as $2 \ \sigma$ RMS.  These particles were tracked with the machine set to the working point $q_h=2.455$ and $q_v=1.415$. This working point gives the most space free of strong resonance lines. Then the tunes were calculated for each surviving particle with PyNAFF. The tune footprint is inserted into FIG. \ref{fig:scanEC}.

Several effects contributes to the tune footprint. The chromaticity is an important one, specially the horizontal.  With the electron cooler included  $\xi_h=-4.4$, $\xi_v=-0.55$ . Another contribution comes from the betatron amplitude dependence of the tunes. The third and probaly the hardest contribution to deal with is the space charge. The effect of space charge is not investigated in this work, but is clearly an important factor to consider. According to~\cite{ElenaDesignRep}, the incoherent tune shift can be -0.1 for a short time before extraction when the beam is bunched. For a coasting beam it is much less, about -0.01, but for a much longer time in the order of seconds. Even in the case of coasting beam the contribution of the space charge to the tune footprint is significant. Therefore it is important to choose a working point which have a sufficiently large space around it in the tune diagram without strong resonance lines, which is the working point $q_h=2.455$ and $q_v=1.415$.
\section{Dynamic aperture}
Frequency analysis allowed to compare and select the promising working points by giving estimates of the relative strength of resonances in the tune diagram. The next step is to analyze the effect of these resonances on the beam. The concept of dynamic aperture quantifies the available stable phase space volume for the beam.
\subsection{The method}
The stability domain is a connected region in phase space of initial conditions that remains bounded after $N$ turns \cite{PhysRevE.53.4067}. 

The stability region is the area inside the last simply-connected invariant curve in phase space. In 4D, the invariant curves do not always separate a connected region, but in most practical cases it exists. We will see later in the results that even if the working point is on a resonance line a connected region can be identified.

In 4D, the volume of the stability domain can be calculated by the integral
\begin{equation}
\int \int \int\ \int \chi (x,p_x,y,p_y) \  dx \  dp_x \ dy \ dp_y , \label{EqstabRegion}
\end{equation}  
where $\chi (x, p_x,y,p_y)$ is a function of initial conditions, with value one if the particle has survived $N$ turns, or zero if it was lost.

The dynamic aperture is defined as the radius of the hyper sphere having the same volume in phase space as the stability domain. In practical calculations the phase space coordinates $(x,p_x,y,p_y)$ of Eq. (\ref{EqstabRegion}) are expressed in terms of polar coordinates $\alpha, r$ in physical space and  the phase space angles $\upsilon_1, \upsilon_2 $, then the integral approximated  as a sum over these coordinates.

The direct calculation of  Eq. (\ref{EqstabRegion}) by summation over all four variables is very  CPU time demanding. Instead we applied the method described  in \cite{Giovannozzi96_NumMethForDA,PhysRevE.53.4067} as `integration over the dynamics'. The idea is to replace the integration over the phase space angles $\upsilon_1, \upsilon_2 $ by averaging over the dynamics. While particles take turns, they sweep the full range of phase space angles $\upsilon_1, \upsilon_2 $, and this information is available. 
 
The procedure is the following.  With fixed initial phase space angles $\bar{\upsilon_1},\bar{\upsilon_2}$, the polar angle $\alpha_k$ is scanned in $K$ steps. For each $\alpha_k$ the  radius  of the last stable orbit $ r(\alpha_k,\bar{\upsilon_1},\bar{\upsilon_2} )$ is sought by increasing the polar radius variable in $J$ steps. Then the $N$ turns of the last stable orbit for each $\alpha_k$ is distributed into an angular grid of size $M \times M$, such that each grid element should contain at least one iterate from the orbit of $ r(\alpha_k,\bar{\upsilon_1},\bar{\upsilon_2} )$. Then the dynamic aperture is calculated with the formula                                                                                                                                                                                                                                                                                                                                                                                                                                                                                                                                                                                                                                                                                                                                                                                  
\begin{equation}
r_{d}= \left[ \frac{\pi}{2 K M^2} \sum_{m1,m2=1}^M \sum_{k=1}^K  r_{m1,m2}(\alpha_k,\bar{\upsilon_1},\bar{\upsilon_2} ) ^4 sin(2 \alpha_k) \right ]^{\frac{1}{4}},
\label{Eqdynap}
\end{equation}  
as given in \cite{PhysRevE.53.4067}.

We found that $M$ could not be always increased above a rather low value without having at least one empty bin in the $M \times M$  grid. This is due to the very inhomogeneous distribution of phase space angles $\upsilon_1, \upsilon_2 $, which is a known phenomenon \cite{RevModPhys.64.795, PhysRevE.53.4067}. A low value for $M$  leads to low accuracy.

There is another way to calculate the dynamic aperture based on normal forms given to the first order in \cite{PhysRevE.53.4067} as
\begin{equation}
r_{NF}= \left[ \frac{\pi}{2 K}  \sum_{k=1}^K  \{\rho_1(\alpha_k,\bar{\upsilon_1},\bar{\upsilon_2} ) + \rho_2(\alpha_k,\bar{\upsilon_1},\bar{\upsilon_2} ) \} ^2 sin(2 \alpha_k) \right ]^{\frac{1}{4}},
\label{normalFormDA}
\end{equation}  
where $\rho_1(\alpha_k,\bar{\upsilon_1},\bar{\upsilon_2} )$ and $ \rho_2(\alpha_k,\bar{\upsilon_1},\bar{\upsilon_2})$ are the nonlinear invariants, which are the same as the Courant-Snyder invariants. Since there is already an implementation of ellipse fit to the phase space variables in SIMPA, the values of $\rho_1(\alpha_k,\bar{\upsilon_1},\bar{\upsilon_2} )$ and $ \rho_2(\alpha_k,\bar{\upsilon_1},\bar{\upsilon_2})$ are easily obtained. It should be noted however, this sum can lead also to non-negligible errors if the working point is on a strong resonance line and the form of the phase space plot is significantly different from an ellipse. Nevertheless, it is still the best first-order approximation of the dynamic aperture. For this reason we calculated with both Eq. (\ref{Eqdynap}) and Eq. (\ref{normalFormDA}) as a cross check.

In our numerical experiments, the closed orbit was different from the design orbit. The electron cooler has introduced an orbit distortion, which was corrected. However, the correction was not perfect. The maximum orbit deviation was less than 3 millimeters horizontally and less than 1 millimeter vertically. It could have been reduced further, but doing so for each point we investigated is rather tedious. Also, in reality, the orbit of ELENA during the commissioning was bigger than that.
\subsection{Results}
\begin{figure}
\includegraphics[scale=0.55]{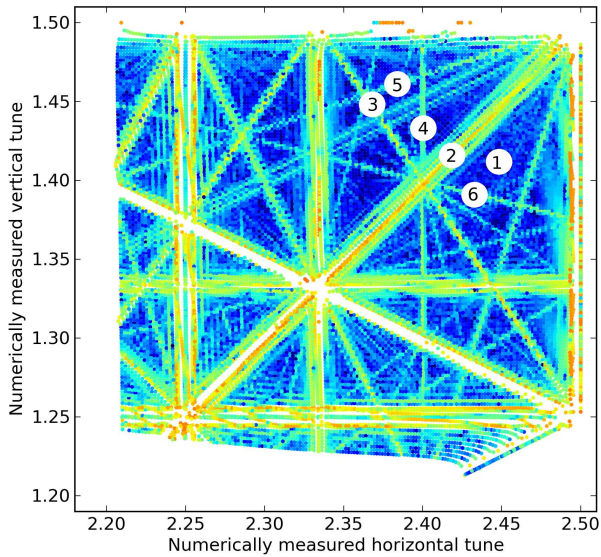}
\caption{\label{fig:points} The numbered white points in the tune diagram indicate the tunes where the dynamic aperture was calculated.} 
\end{figure}

The dynamic aperture of six working points has been evaluated to compare resonance conditions with a non-resonant case. The selected set of tunes is indicated on FIG. \ref{fig:points}. These resonance lines were selected because they are close to the region where the working point of ELENA was set during the commissioning. The first working point was chosen to be far from strong resonance lines at $Q_h=2.455, \ Q_v=1.415$. The other five points were placed on resonance lines with various orders.  The dynamic apertures calculated with Eq. (\ref{Eqdynap}) and  Eq. (\ref{normalFormDA}) are displayed in TABLE \ref{tableDA}. All the results were calculated with $N=10^4$ turns,  $K=20$ polar steps, with  $\alpha_k \in [0,\pi/2 ]$, radial steps $J=20$ and the biggest possible $M$, which is also indicated in TABLE \ref{tableDA}. The energy of the particles was set to 100 keV. We also calculated  the errors of the dynamic aperture values in in TABLE \ref{tableDA} with Eq. (10) given in \cite{Giovannozzi:1997uc}, it is $\pm 0.0011 [m]$.

\begin{table}
\caption{Dynamic apertures for the six points in FIG. \ref{fig:points} calculated with averaging over the dynamics ( $r_d$) and with normal forms  ($r_{NF}$). M is the angular grid size for  $r_d$ calculation.
For the tunes falling onto a resonance line, the equation of the resonance is also given.\label{tableDA}}
 \begin{center}
 \begin{tabular}{c c c c c } 
 \hline
 Point  & \ \ $r_d \ [m]$  &\  M &\ \  $r_{NF} \ [m]$ & \ \ Resonance condition \\ [0.5ex] 
 \hline\hline
 1 &  0.0115 &14 &0.0115 & NA \\ 
 \hline
 2 & 0.012 &8 &  0.0112& $Q_h-Q_v = 1$ \\
 \hline
 3 &  0.01 &6& 0.01 & $3 Q_h+2 Q_v= 10$ \\
 \hline
 4 & 0.0117 & 3 & 0.0114 &  $5 Q_h = 12$ \\
 \hline
 5 & 0.0113 & 9& 0.011 & $-Q_h+3 Q_v=2$\\
 \hline
 6 & 0.0113 &8 & 0.0108 & $Q_h+4 Q_v=8$\\ [1ex] 
 \hline
 \hline
\end{tabular}
\end{center}
\end{table}
\begin{figure}
\includegraphics[scale=0.88]{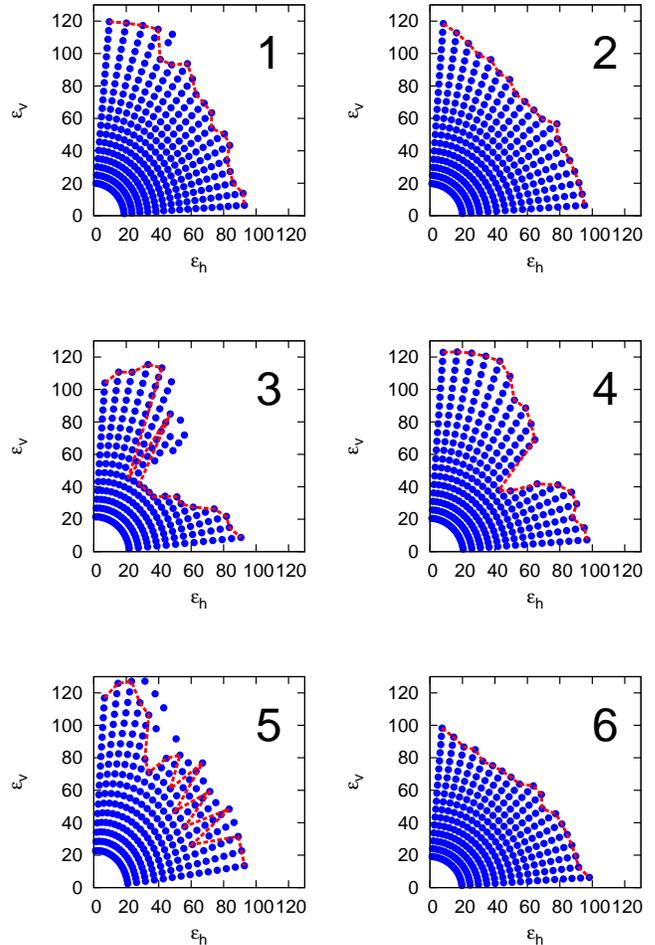}
\caption{\label{fig:multiDA} Initial emittances plotted in units of $[\pi \ mm \ mrad] $ for particle with $dp/p=0$. The horizontal axis corresponds to the horizontal initial emittance and the vertical axis to the initial vertical emittance. Only those initial conditions are plotted which survived $N=10^4$ turns. The numbers in the upper right corners correspond to the numbering in FIG.  \ref{fig:points}. The lines indicate the last connected initial emittances for each angle $\alpha_k $.} 
\end{figure} 
\begin{figure}
\includegraphics[scale=0.88]{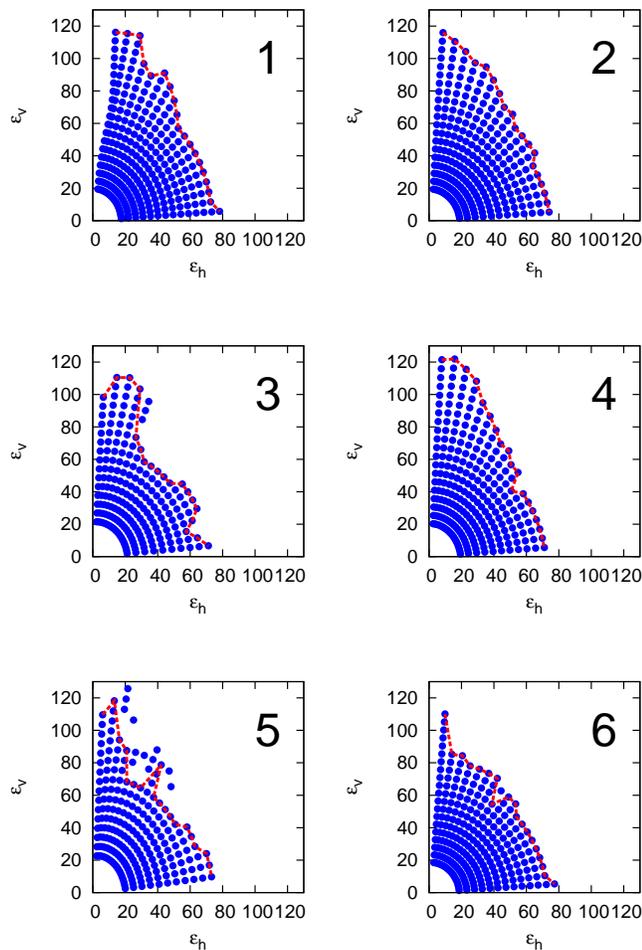}
\caption{\label{fig:multiDAdp} Initial emittances plotted in units of $[\pi \ mm \ mrad] $ for particle with $dp/p=2 \times 10^{-3}$.}
\end{figure} 
Most of the dynamics aperture values in TABLE \ref{tableDA} are close to 0.011. At first sight one is tempted to conclude that the resonance lines of order 4 or 5 are already weak enough to not reduce further the physical aperture of the machine. The dynamic aperture however is defined as the radius of a four dimensional hyper sphere and small differences in the radius can hide significant phase space volume reductions at specific angles of $\alpha_k$.

For this reason we found it informative to give also a more detailed description by plotting those initial emittances in FIG. \ref{fig:multiDA} for each point in FIG. \ref{fig:points}, which survived $N=10^4$ turns. We repeated the same exercise with particles having a momentum deviation $dp/p=2 \times 10^{-3}$, which is the maximum momentum spread measured during the commissioning. The quadrupole currents were adjusted for each point to have the same tunes as for the case  $dp/p=0$. The results are plotted in FIG. \ref{fig:multiDAdp} .

In FIG. \ref{fig:multiDA}, \ref{fig:multiDAdp} it is apparent that some of the resonance lines make the dynamics aperture significantly smaller than the physical aperture.
The physical acceptance of ELENA is determined by the  size of vacuum chambers and the optics.

There are a number of interesting features displayed in these figures, some of them surpising. The coupling resonance $Q_h-Q_v = 1$ indicated by point two is rather wide and looks strong, but it does not significantly reduce the dynamic aperture. The smallest dynamic aperture was obtained at  point three, due to a 5th order resonance and not a lower order resonance line. We did not include first, second and third order lines, those should obviusly be avoided during operation. Higher-order resonance lines however can not always be evaded. The incoherent space charge tune shift in ELENA at the lowest energy is significant. Its estimated value for a bunched beam at arrival to 100 KeV is  $\Delta Q= -0.03$  \cite{ElenaDesignRep}. The incoherent space charge tune shift contributes to the tune footprint of the beam and some of the resonances shown in FIG. \ref{fig:points} are unavoidable. Therefore it is important to choose a working point which gives the most space free of harmful resonance lines. Point one in FIG. \ref{fig:points} is a good choice, because the surrounding resonance lines  $Q_h-Q_v = 1$ (on point two) and $Q_h+4 Q_v=8$ (on point six) are rather forgiving according to FIG. \ref{fig:multiDA}, \ref{fig:multiDAdp}.

The initial emittances for the test particle to be tracked were calculated by a linear model, that is the Courant-Snyder theory. It assumes that the values of the phase space variables lie on an ellipse. This is only approximately true under resonance conditions. Although the angle $\alpha_k$ was changed linearly in equal steps, in FIG. \ref{fig:multiDA} case 5 for instance, shows distortions. This is because the phase space plot deviates significantly from an ellipse and the requested emittance is different from the actual initial emittance, which is plotted. 

\begin{figure}
\includegraphics[scale=0.9]{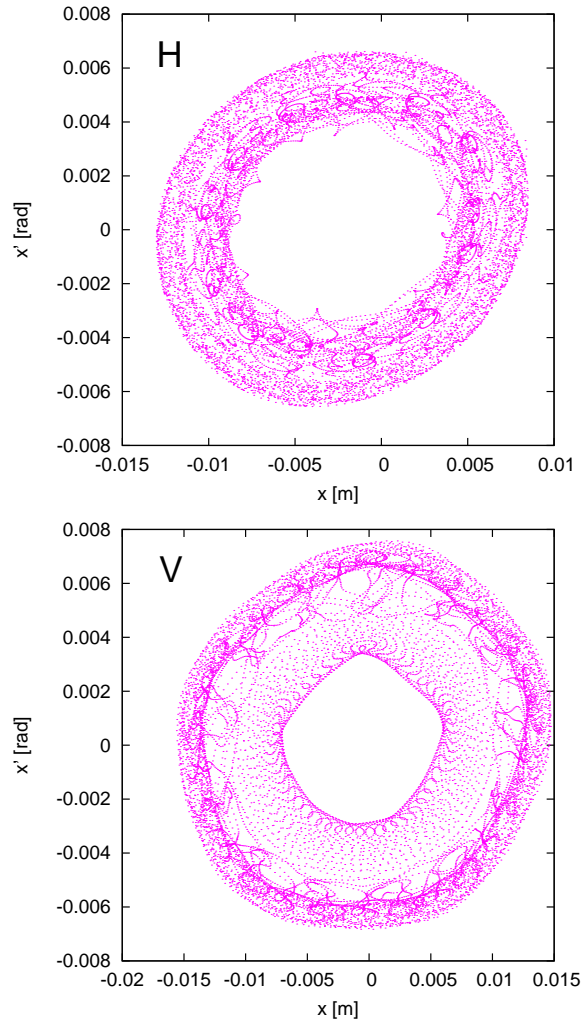}
\caption{\label{fig:phs} Phase space plots of a particle near a 4th order resonance, lost after 14211 turns.} 
\end{figure}

The dynamic aperture is defined as a function of turns. It is known to decrease with the inverse of the logarithm of the number of turns \cite{Giovannozzi96_BDN,Giovannozzi:1996PredLong,Giovannozzi:1997uc}. Indeed it is easy to find initial conditions which leads to particle loss after more than $10^4$ turns, an example is given in FIG. \ref{fig:phs}.
\section{Conclusions}
We have applied the SIMPA code for long-term symplectic charged particle tracking in arbitrary static electromagnetic fields on the ELENA machine. The algorithm described in a previous paper has been extended to handle complex beam region geometries. A short description of the enhanced algorithm was given. The improvements not only raised its generality, but also increased the speed by a factor 4.66. 

The frequency and dynamic aperture analysis identified a number of 4th and 5th order resonance lines in the tune diagram strong enough to reduce the dynamic aperture below the physical one. What made the frequency and dynamic aperture analyses different, is the fact that we have not introduced any multipole error into our model apart from the multipole components due to the geometry of the magnets, which are inevitable. All the resonances seen in the frequency analysis are direct consequences of the geometry of the fields, even if the magnetic elements are manufactured perfectly. 

We showed in the frequency analysis section the effect of the magnetic system of the electron cooler on the beam dynamics by comparing the two cases, with and without electron cooler. The electron cooler introduced non-negligible magnetic perturbations, strengthening many resonance lines. To our best knowledge there was no similar study done before with electron coolers.
\subsection{Suggestions to improve the performance}
The commissioning of the ELENA ring was finished in November 2018, at the start of the 2.5 years long CERN wide shutdown. The deceleration efficiency was estimated between 40 - 50 \%, somewhat below the $60 \ \%$ design value. Most of the losses occurred on the ramp just before the lowest energy plateau. This is not surprising, since both the incoherent tune shift due to space charge and the intrabeam scattering has the biggest effects at low energy. Also the perturbation introduced by the electron cooler is the strongest at the lowest energy. 

During the commissioning two working points were explored extensively, $Q_h=2.42,\ Q_v=1.45$, and $Q_h=2.38,\ Q_v=1.44$. These working points are not far from the 4th and 5th order resonances number 3, 4,  5 in FIG. \ref{fig:points} and these resonance lines are not harmless according to FIG. \ref{fig:multiDA}, \ref{fig:multiDAdp}. The tune footprint of the beam certainly overlaps with some of the resonance lines. The incoherent tune shift of a bunched beam due to space charge at arrival to 100 KeV energy is in the order of -0.03, depending on the beam intensity and beam size. This is not a negligible value, but we believe it is not big enough to drastically change the results of our analysis. One supporting evidence for this assumption, is that we haven't observed significant dependence of the deceleration efficiency  on the beam intensity during the commissioning. After cooling at 100 keV energy, the incoherent tune shift is in the order of -0.1, but only for the duration of the final part of the bunching before extraction, about 10 milliseconds. No significant losses were observed related to the bunching process during the commissioning. Based on the frequency and dynamic aperture analysis, we suggest to explore working point number one in  FIG. \ref{fig:points}, as it gives the most available space without harmful resonances.  

The tune footprint of the beam with maximum emittances in all planes shown on FIG. \ref{fig:scanEC} is mostly coming from the horizontal chromaticity which is $\xi_h=-4.4$ at 100KeV energy with electron cooler included. Compensating the chromaticity reduces the tune spread and can help to avoid resonances. There are sextupoles foreseen in ELENA for this, but during the commissioning they were not used. We believe it to be worth trying.

\section{Acknowledgments}

I would like to express my gratitude towards Gianluigi Arduini, Christian Carli and  Massimo Giovannozzi from CERN BE-ABP  for their valuable comments and suggestions. 
\bibliography{dynap}   

\end{document}